\documentclass[10pt,twocolumn,amsmath]{article}

\usepackage{graphicx}
\usepackage{bm}
\usepackage{latex8}
\usepackage{times}

\begin{document}

\title{Simulations of Disordered Bosons on Hyper-Cubic Lattices}

\author{Peter Hitchcock and Erik S. S\o rensen\\
Department of Physics and Astronomy, McMaster University, Hamilton, ON, L8S 4M1 Canada\\
hitchpa@mcmaster.ca, sorensen@mcmaster.ca}

\date{\today}

\maketitle

\begin{abstract}
We address computational issues relevant to the study of disordered quantum mechanical
systems at very low temperatures. As an example we consider the disordered Bose-Hubbard
model in three dimensions directly at the Bose-glass to superfluid phase transition. The universal
aspects of the critical behaviour are captured by a (3~+~1)~dimensional link-current model for which
an efficient `worm' algorithm is known. We present a calculation of the distribution of the
superfluid stiffness over the disorder realizations, outline a number of important considerations
for performing such estimates, and suggest a modification of the link-current Hamiltonian
that improves the numerical efficiency of the averaging procedure without changing the universal
properties of the model.
\end{abstract}

\Section{Introduction}
The Bose-Hubbard model was first studied in the context of liquid helium in a disordered
medium~\cite{fisher:1989:prb}. Interest in the model has recently grown with the progress achieved in trapping
ultra-cold atomic gases in optical lattice potentials. The model describes the competition between
tunneling and on-site interactions in a lattice of bosons. It displays several zero-temperature
quantum phases that are now clearly attainable in the laboratory~\cite{greiner:2002a:nature}.
Notably, a localized Mott insulating (MI) phase exists when the tunneling between sites is small,
while at higher tunneling the system becomes a superfluid (SF). In the presence of disorder another
localized phase, the Bose-glass (BG), exists between the Mott insulator and the superfluid. In the
present study we focus on the Bose-glass to superfluid transition since it exists only at finite
disorder and thus provides a clear example of a quantum phase transition for which disorder is
relevant. Although this model has been extensively studied in one- and two-dimensions, the nature of
the phase transition in three and higher dimensions has received relatively little attention.

Scaling theories based on generalized Josephson relations and the finite compressibility of the
superfluid and Bose-glass phases indicate that the dynamic correlation exponent $z$ is equal to the
number of spatial dimensions~\cite{fisher:1989:prb}. This feature, which is supported by analytical
and numerical arguments~\cite{herbut:1998:prb,herbut:2001:prl,wallin:1994:prb} in low dimensions
suggests that the model has an unusual approach to mean-field behaviour---and may not have an upper
critical dimension at all---invalidating standard renormalization group approaches. Numerical work
above two dimensions is difficult due to the large volumes of the systems, and the 
algorithmic slow down of the Monte Carlo averaging procedure. 

Of particular interest are the distributions of thermodynamic observables over the disorder. They
are typically far from Gaussian in nature, rendering standard estimates of statistical error
invalid for smaller sample sizes and necessitating calculations for a large number of disorder
configurations. Moreover, the behaviour of these distributions for increasing lattice size is 
directly relevant to the break-down of self-averaging~\cite{aharony:1998:prl}, the quantum Harris
Criterion~\cite{chayes:1986:prl,pazmandi:1997:prl}, and the effect of disorder on quantum critical
phenomena~\cite{sknepnek:2004:prb}. More efficient ways of performing disorder averages has also
been proposed~\cite{bernadet:2000:prl}. These latter developments are however too computationally demanding for
the present model.

The universal properties of the $d$-dimensional Bose-Hubbard model are captured by a
($d$~+~1)-dimensional classical link-current representation for which an efficient worm-like Monte
Carlo algorithm exists. While the worm algorithm represents a drastic improvement over earlier,
Metropolis-like algorithms, the computational demands increase dramatically in higher dimensions,
limiting the precision of the numerical analysis.  Since the system is disordered, the calculations
involve performing many Monte Carlo simulations of the system at the same parameters with different
realizations of the disorder. Hence, the calculations are very well suited for parallelization and
a linear speed up can be achieved with a relatively modest programming effort. Without such a linear
speed up the calculations we report on would have been almost impossible.  Parallelization is
performed straightforwardly with MPI: each processor performs Monte Carlo simulations serially for
a given disorder realization, then the results are collected and written to disk. As many as several
thousand disorder realizations need to be performed at each point in parameter space, and each
individual Monte Carlo simulation can take up to three hours, depending on the
system size and the parameters of the model. The length of time required to perform one simulation
for a given disorder realization dictates how large a system we can reasonably study and so it is
essential to consider carefully how much computational effort to invest in each such simulation.

Numerical study of this transition in three dimensions presented a number of difficulties in
calculating the disorder distributions and their averages. The main results of the study will be
presented elsewhere~\cite{hitchcock:2006:prb}, but we outline here the procedure that was used to
estimate these distributions, and suggest a modification to the link-current Hamiltonian which
improves the efficiency of these numerical estimates. The remainder of the introduction discusses
the link-current Hamiltonian and the finite-size scaling theory on which our numerical approach
relies.  Section~\ref{sec:eq} describes how we ensure that the simulation of each disorder
realization has been properly equilibrated. Section~\ref{sec:ar} describes a modification of the
link-current model that improves the numerical efficiency of each simulation without affecting the
universal details. We then conclude with some remarks about the general applicability of these
considerations.

\SubSection{Model and Scaling Theory}
The Bose-Hubbard Hamiltonian, including an on-site disorder in the chemical
potential is~\cite{fisher:1989:prb}
\begin{equation} 
H_{\mathrm{BH}} = \sum_{{\bf r}}\Big[\frac{U}{2}\hat{n}_{{\bf r}}(\hat{n}_{{\bf r}} - 1) - \mu_{{\bf r}}\hat{n}_{{\bf r}}\Big] - 
\frac{t}{2}\sum_{\langle {\bf r}, {\bf r}'\rangle}(\bm{\hat{\Phi}^\dagger_r\hat{\Phi}_{r'}} + \mathrm{H.c.}).  
\end{equation}
The second quantized boson operators describe a tunneling process coupled by $t$ and an on-site,
repulsive interaction $U$, on a hyper-cubic lattice. The disordered chemical potential $\mu_{\bf r}$
is distributed uniformly on $[\mu - \Delta, \mu + \Delta]$ so that $\Delta$ controls the strength of the
disorder. At finite disorder, the system undergoes a phase transition from a Bose-glass insulating
phase (low $t$, high $U$) to a superfluid phase (high $t$, low $U$). The model can be transformed via the
quantum rotor model to the ($d$~+~1)~link-current model~\cite{wallin:1994:prb}. The link-current
Hamiltonian is given by 
\begin{equation} 
\label{eq:Hlc} 
H = \frac{1}{K}\sum_{({\bf r}, \tau)}\Big[\frac{1}{2}\bm{J}^2_{({\bf r}, \tau)} - \mu_{{\bf r}}J^\tau_{({\bf r}, \tau)}\Big].  
\end{equation}
The integer currents ${\bf J}_{({\bf r}, \tau)}$ are situated on the bonds of the lattice and obey a
divergenceless constraint 
\begin{equation}
\sum_{\nu = x, y, z, \tau} J^\nu_{({\bf r}, \tau)} = 0. 
\end{equation}
The resulting loops are interpreted as currents of bosons hopping about on the lattice (specifically
they are fluctuations from an average density, so that the currents are permitted to be negative). The
coupling $K$ controls the ratio between $t$ and $U$: at low $K$ the interaction $U$ dominates and the system
is insulating, while at high $K$ the tunneling dominates and the system condenses into a superfluid. 

The two phases can be distinguished by the superfluid stiffness, $\rho$, which is proportional to the
superfluid density. The stiffness is defined as the response of the free energy to a twist in the
boundary conditions and is indicative of global phase coherence. In the link-current model, the
stiffness is proportional to the square of the winding number in the spatial dimensions: 
\begin{equation}
\label{eq:stiffness}
\rho = \frac{1}{L^{d - 2}L_\tau}\big[\langle n_{x}^2 \rangle\big]_{av}.
\end{equation}
The angle brackets $\langle\cdot\rangle$ denote a thermal average, while the square brackets
$[\cdot ]_{av}$ denote an average over disorder realizations. The winding numbers ($n_{\gamma} =
L_\gamma^{-1}\sum_{{\bf r}, \tau}J^{\gamma}_{{\bf r}, \tau}$ for $\gamma = x, y, z, \tau$) of the
lattice in each direction are just the number of current loops that have wound all the way about the
periodic lattice. They are always integers.

Dynamics and statics are both essential to the critical behaviour of quantum phase transitions. They are 
characterized by independent spatial and temporal correlation lengths ($\xi$ and $\xi_\tau$) which
define the dynamic critical exponent $z$:
\begin{equation} 
\xi_\tau \sim \xi^z \sim (\delta^{-\nu})^z,\qquad\delta = \frac{K - K_c}{K_c}.
\end{equation} 
Here $\nu$ is the correlation length exponent. This implies that quantities at the critical point
scale as a function of two arguments. The superfluid stiffness (which diverges as $\rho \sim \xi^{d
+ z - 2}$ at $K_c$) thus scales as
\begin{equation}
\label{eq:pscale}
\rho = \frac{1}{L^{d + z - 2}}\bar{\rho}(L^{1/\nu}\delta, L_\tau/L^z).
\end{equation}
If we hold the second argument constant, the critical point can be located by
plotting curves of $\rho L^{d + z - 2}$ for various linear system sizes $L$. 
Since $\delta = 0$ at $K_c$, 
$K_c$ will be  the value of $K$ at
which these curves intersect. This unfortunately requires that we guess at the value of $z$ before
we begin. In principle, we are free to set the aspect ratio
\begin{equation}
\alpha = L_\tau/L^z,
\end{equation}
as we see fit
to find the critical point; in practice however, as we discuss below, the numerics work better near
an optimal aspect ratio where $(\xi/L)^z\simeq\xi_\tau/L_\tau$ implying $\alpha = {\cal O}(1)$.  

\Section{\label{sec:eq}Equilibration}
When considering disordered systems, the average $[\langle\cdot\rangle]_{av}$ of an arbitrary
observable (denoted by $\cdot$) such as the stiffness $\rho$ must be calculated over a whole set of disorder
configurations, performing independent Monte Carlo simulations on each particular realization. We
must then decide how many Monte Carlo sweeps ($t_n$) to perform on each simulation, and how many
disorder realizations ($N_D$) to average these over.  There are correspondingly two sources of
statistical error~\cite{bhatt:1988:prb}: the error $\delta_T\rho$ in the estimate of the thermal
average: 
\begin{equation}
\langle\rho\rangle = \bar{\rho} + \delta_T\rho,
\end{equation}
and the overall error $\delta[\rho]_{av}$ in the disorder average:
\begin{equation}
[\langle\rho\rangle]_{av} = [\bar{\rho} + \delta_T\rho]_av = [\bar{\rho}]_{av} + \delta[\rho]_{av}.
\end{equation}
It is of particular interest to review~\cite{bhatt:1988:prb} how to correctly obtains the disorder average of
the square of a thermodynamic observable such as the energy. Such a quantity would be needed
for calculating for instance the specific heat, $C_V$. In this case we have for
a single disorder realization:
\begin{equation}
\langle E\rangle = \bar{E} + \delta_T E.
\end{equation}
If we now want to calculate the disorder average $[\langle E\rangle^2]_{av}$ we encounter a
slight problem:
\begin{equation}
[\langle E\rangle^2]_{av} = [\bar{E}^2]_{av} + [(\delta_T E)^2]_{av} + 2\bar{E}[\delta_T E]_{av}.
\end{equation}
It is natural to assume that $[\delta_T E]_{av}$ will yield zero when a sufficiently large number
of disorder realizations are used. However, $[(\delta_T E)^2]_{av}$ will be {\it non zero} and will
yield a {\it systematic} error unless infinitely precise thermal averages can be obtained for each disorder realization.
In order to circumvent this problem and correctly calculate such a disorder average,
one can 
run 2 independent simulations, referred to as ``replicas'', of a given disorder realization~\cite{bhatt:1988:prb}.
We
denote them by $\alpha$ and $\beta$.  The above disorder average should then be calculated in the
following way: 
\begin{eqnarray}
[\langle E\rangle^2]_{av} &\equiv& [\langle E^\alpha\rangle\langle E^\beta\rangle]_{av} \nonumber\\
  &=&[\bar{E}^2]_{av} +
[\delta_T E^\alpha\delta_TE^\beta]_{av}\nonumber\\
  & &+\bar{E}[\delta_T E^\alpha]_{av}+\bar{E}[\delta_T E^\beta]_{av}.
\end{eqnarray}
The term $[\delta_T E^\alpha\delta_TE^\beta]_{av}$ will now also correctly average to zero since the thermal
errors from each replica are independent random variables.
In our calculations we always run at least two independent replicas of a given disorder realization
with the goal of correctly calculating averages as outlined above. For higher powers of thermal averages
more replicas are needed. As we shall see below, having several independent runs of a given disorder realization
allow also for very convenient and indispensable checks of the equilibration of the calculations.

\begin{figure}[t]
\includegraphics[clip,width=8cm]{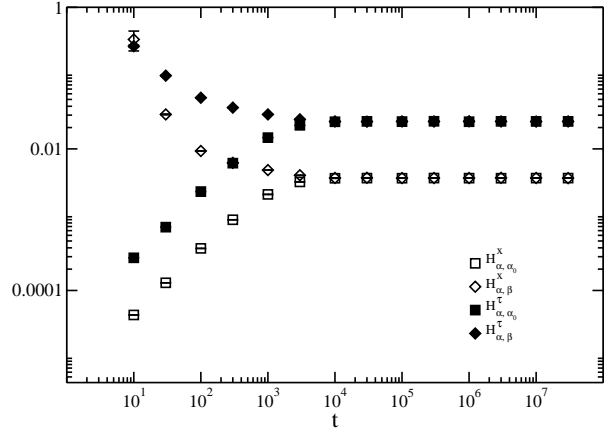}
\caption{ Hamming distances on an 8x8x8x64 lattice at ${\bf K_c = 0.19}$ calculated over a set of
				1000 disorder realizations with ${\bf t_0 = 3\times 10^7}$. For the Hamming distances between the
				two replicas ${\bf \alpha}$ and ${\bf \beta}$, ${\bf t}$ is the total number of Monte Carlo sweeps
				performed. For the Hamming distances between replica ${\bf \alpha}$ and its configuration
				${\bf \alpha_0}$ at ${\bf t_0}$, ${\bf t}$ is the number of sweeps performed {\it after}
				the initial ${\bf t_0}$ sweeps have been performed. The convergence of the curves indicates 
				${\bf t_r \approx 3\times 10^5}$.} 
\label{fig:hamm}
\end{figure}

It is important to note that even for very large system sizes, the average should be taken over
as many disorder realizations as possible.  One might assume that for large system sizes a
smaller set of disorder realizations would suffice, since the properties of the system will average
out spatially. For many disordered systems this assumption is false---self-averaging breaks
down~\cite{aharony:1996:prl}.  Even in the thermodynamic limit (where in principle one could find
any particular finite disorder realization \emph{somewhere} in the infinite system) the width of
the distribution of $P(\bar{\rho})$ remains finite.  Moreover, such distributions are typically far
from Gaussian (they often have a particularly `fat' tail). Error estimates based on Gaussian
distributions are thus only valid for very large $N_D$.  The best approach is then to spend a minimal
amount of computational time on each realization, and then rely on the disorder average to control the
statistical errors. As usual, however, each Monte Carlo simulation must be properly equilibrated to
ensure that $\langle\rho\rangle$ is unbiased. Since the lattice starts in an artificial (and
non-equilibrium) configuration, we throw out the first $t_0$ of the $t_n$ sweeps before sampling the
remaining $t_s = t_n - t_0$ configurations. As long as $t_0$ is greater than the relaxation time of
the algorithm $t_r$, we sample only equilibrium configurations, and our estimates of the thermal
averages will be unbiased. Since we do not want to spend all of our computational efforts
equilibrating systems, each disorder realization is typically run for an additional $t_0$ steps once
it has reached equilibrium. 

To confirm that we have chosen $t_0$ greater than $t_r$, we perform each simulation independently on two 
replicas with different initial configurations.  We can then define
`Hamming distances'~\cite{wallin:1994:prb} between the two replicas $\alpha$ and $\beta$ after
performing $t$ Monte Carlo sweeps on their initial configuration, and between the replica $\alpha$
at sweep $t + t_0$ and its configuration $\alpha_0$ at sweep $t_0$ when sampling begins:
\begin{displaymath}
\nonumber 
H_{\alpha, \alpha_0}^{\nu = x, \tau}(t) = \frac{1}{L^dL_\tau} \sum_{({\bf r}, \tau)}
\big[J^\nu_{\alpha:({\bf r}, \tau)}(t + t_0) - J^\nu_{\alpha:({\bf r}, \tau)}(t_0)\big]^2,
\end{displaymath}
\begin{equation}
\label{eq:hamming}
H_{\alpha, \beta}^{\nu = x, \tau}(t) = \frac{1}{L^dL_\tau}\sum_{({\bf r}, \tau)}
\big[J^\nu_{\alpha:({\bf r}, \tau)}(t) - J^\nu_{\beta:({\bf r}, \tau)}(t)\big]^2.
\end{equation}
These measures are then averaged over the disorder realizations. At the beginning of each
simulation, the initial configurations of $\alpha$ and $\beta$ are quite different. Hence
$H_{\alpha, \beta}(t)$ will be large for small $t$, but diminish as the two configurations
equilibrate. On the other hand, shortly after $t_0$, the configuration $\alpha$ will not have
changed substantially from $\alpha_0$, so $H_{\alpha, \alpha_0}(t)$ will be initially small, but
increase as more sweeps are performed. If $t_0$ has in fact been chosen greater than $t_r$, these
two measures will converge on the same value in $t_r$ sweeps at which point the 
configurations of $\alpha$, $\alpha_0$, and $\beta$ will be independently drawn from the same
population of equilibrium configurations. Figure~\ref{fig:hamm} shows the two Hamming distances
plotted as a function of $t$ at $K_c$, averaged over $N_D = 1000$ disorder realizations. Results are
shown for Hamming distances defined in terms of spatial ($x,y,z$) currents (open symbols) and the
temporal ($\tau$) current (filled symbols).
Both the spatial and temporal Hamming distances converge, indicating that $t_r \approx 3\times 10^5$
for an 8x8x8x64 lattice at the critical point. 

\begin{figure}
\includegraphics[clip,width=8cm]{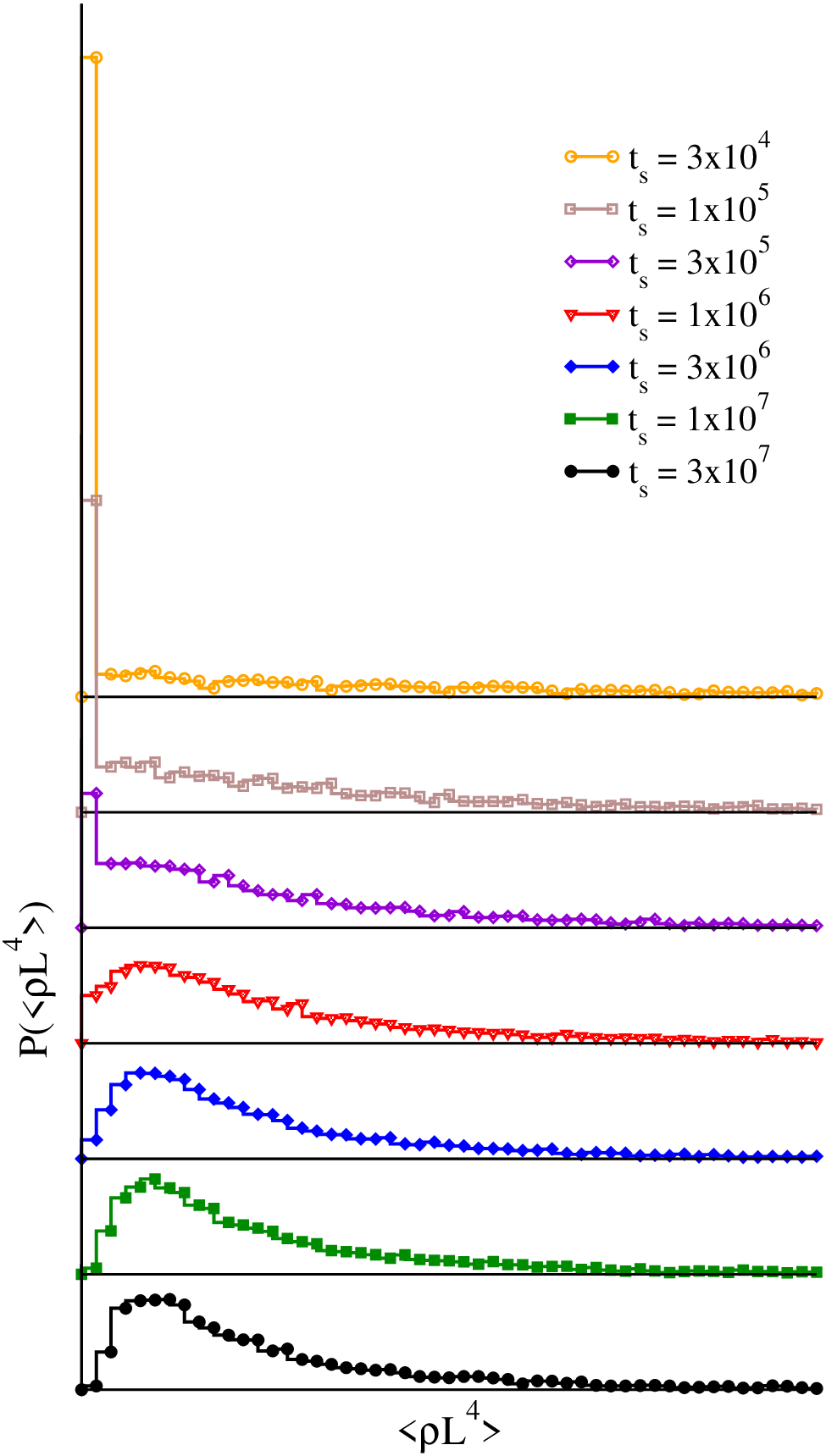}
\caption{ Evolution of the distribution of $\bm{\langle\rho L^4\rangle}$ as a function of $\bm{t_s}$ (the
				number of samples gathered at equilibrium for each disorder realization) for the same sample 
				disorder realizations used in Fig.~\ref{fig:hamm}. The vertical axes are offset for
				clarity. The peak at $\bm{\langle\rho L^4\rangle = 0}$ persists for $\bm{t_s \gg t_r}$, giving 
				way to a broader peak at a non-zero	$\bm{\langle\rho L^4\rangle}$. The distribution continues 
				to change until $\bm{t_s \approx 3\times 10^6}$ upon which the distribution stops evolving.} 
\label{fig:equil}
\end{figure}
\begin{figure}
\includegraphics[clip,width=8cm]{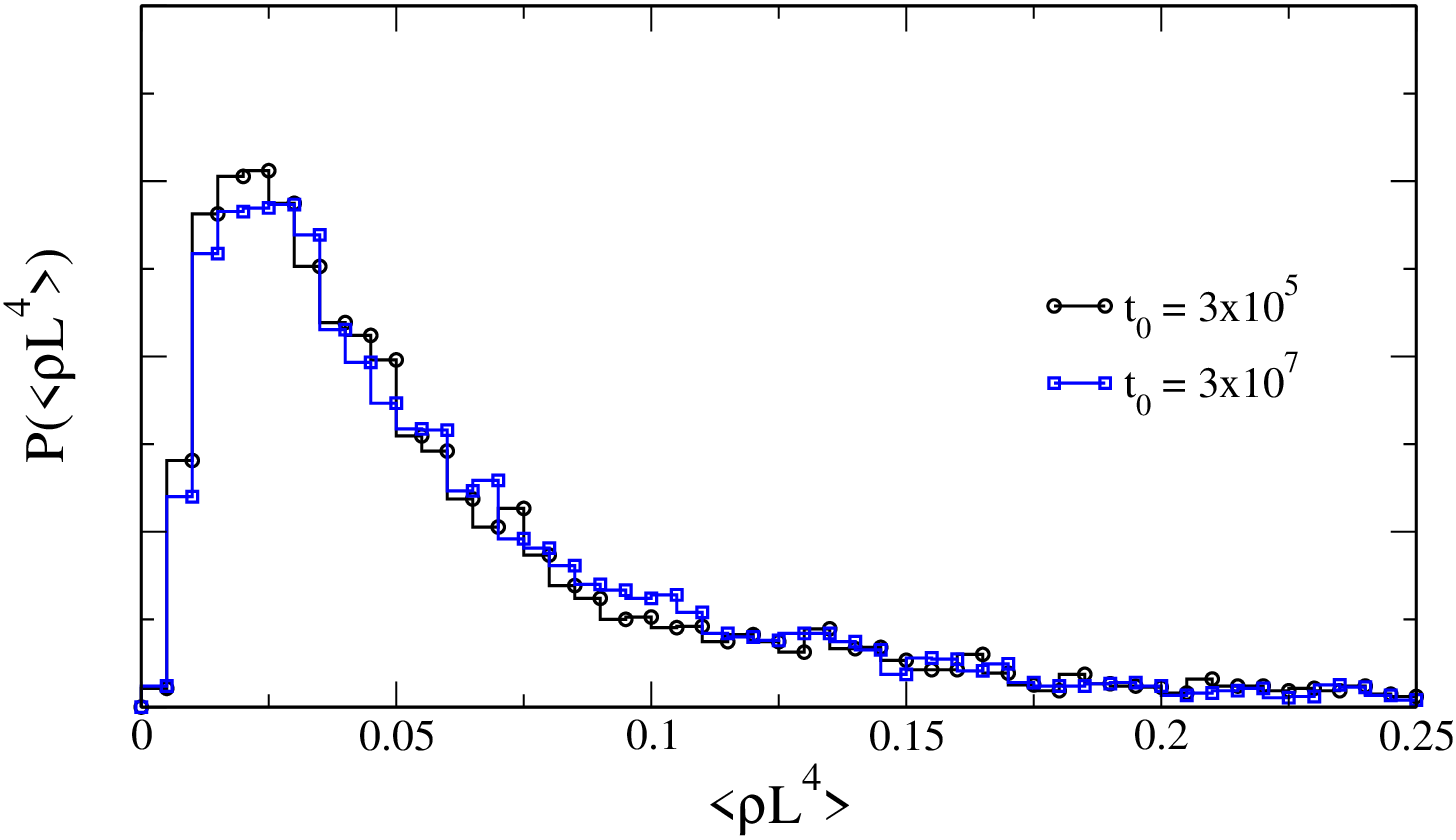}
\caption{ Distribution of $\bm{\langle\rho L^4\rangle}$ calculated with different ${\bf t_0}$ but the same
				$\bf{t_s}$ over two different sets of 1000 disorder realizations. The distribution is unchanged.  }
\label{fig:lowt0}
\end{figure}
 
From these observations it would seem reasonable to generate $t_s = 3\times 10^5$ configurations after equilibration
at each disorder realization in order to calculate the disorder average. However, if we look more closely
at the distribution $P(\langle\rho\rangle)$ generated by using $t_s = 3\times 10^5$ configurations,
we find a large peak in the distribution at $\langle\rho\rangle = 0$ (see the uppermost graph of
Fig.~\ref{fig:equil}). For many disorder realizations, the Monte Carlo algorithm {\it never}
generates an equilibrium configuration with a non-zero winding number (this can be verified by
looking directly at the data set).  If we generate more configurations at equilibrium for each
disorder realization (that is, we increase $t_s$), the shape of the distribution changes---the peak
at $\langle\rho\rangle = 0$ shrinks and a broader one grows at a non-zero value. After one to three
million sweeps, the peak at $\langle\rho\rangle = 0$ disappears and the distribution stops evolving.
This effect is a result of the discrete nature of the winding number, $n_x$, and the strong
correlation between configurations generated by a successive Monte Carlo sweeps.  Since the
superfluid stiffness is defined in terms of
$n_x$, it is difficult to numerically resolve the difference between a superfluid stiffness of zero
and a small fractional value. For instance, a disorder realization with $\bar{\rho} L^4 = 0.05$
corresponds to an average squared winding number of $\langle\bar{n_x^2}\rangle = \frac{L_\tau}{L^z}\langle\rho
L^4\rangle = 1/160$, roughly implying that only one out of every 160 independent configurations has a
non-zero winding number.  Since the algorithm generates approximately one fully independent configuration every
$t_r$ sweeps, at $t_s = 3\times 10^5$ one would expect most runs to find $\langle\rho
L^4\rangle = 0$. (Here we assume that $t_r$ is proportional to the autocorrelation time $\tau$.)
However, if the average is taken over a further $10^2$ independent configurations, the
estimate will converge on the small but finite true average.  Resolving the true shape of
the distribution thus requires much longer runs at each realization of the disorder. That the
lattice is in fact equilibrated at $t_r$ as determined by the convergence of the Hamming distances
is supported by the fact one can generate the same distributions independent of $t_0$ (so long as
$t_0 > t_r$).  Figure~\ref{fig:lowt0} shows the distribution of $\langle\rho L^4\rangle$ calculated
over two different sets of 1000 disorder realizations, one with $t_0 = 3\times 10^5$, the other with
$t_0 = 3\times 10^7$. In both cases $t_s = 3\times 10^7$ configurations were generated after equilibration
for each disorder realization.

An immediate conclusion to be drawn from these results is that, if attention is not paid to
either the Hamming distances or the convergence of the complete distribution of the thermodynamic
observables over the disorder, then one is very likely to be mislead by the results obtained. In
particular, it is clear from the results in Figure~\ref{fig:equil} that if $t_s$ or $N_D$ are too
small, then $[\langle\cdot\rangle]_{av}$ will be {\it too small} and the error bars will also be
misleadingly small. As $t_s$ and $N_D$ are increased, $[\langle\cdot\rangle]_{av}$ will also likely
increase due to the tails of the distribution while at the same time the associated error bars might
very well remain roughly constant.

\Section{\label{sec:ar}Coupling Anisotropy}
In the context of the Bose-glass to superfluid transition, one approach to improving the efficiency
of the calculation is to increase the numerical value of the stiffness at the critical point.  One
possible means of achieving this is to increase the aspect ratio $\alpha$ between the spatial and
temporal lattice sizes.  In a finite system, the correlation length can be no longer than the size
of the system; but since the temporal and spatial correlation length are related, a small $\alpha$
that truncates $\xi_\tau$ will in turn restrict $\xi$ to be smaller than the spatial extent of the
lattice. Since a global current loop depends on correlation across the whole length of the lattice,
this will limit the number of winding events and in turn reduce $\rho$. At the critical point, the
value of $[\langle\rho L^4\rangle]_{av}$ should increase with $\alpha$ and it seems natural to
assume that it will continue to increase at least up until $\xi_\tau/L_\tau\simeq(\xi/L)^z$, where
$\alpha = {\cal O}(1)$.  Simulations are then best performed at this optimal aspect ratio;
unfortunately they prove too large for us to simulate. The problem here is the relatively small
probability of generating configurations with non-zero winding number in the spatial direction if a
very small $\alpha$ is used.  One could then ask if equivalent models can be found with higher
winding numbers at $K_c$ for the same aspect ratio. Such models would likely have an optimal
$\alpha$ at smaller values than the present one thereby decreasing the numerical overhead. It turns
out that for the present model we can exploit the universality of the critical behavior to arrive at
equivalent models that indeed have this desired property.

\begin{figure}
\includegraphics[clip,width=8cm]{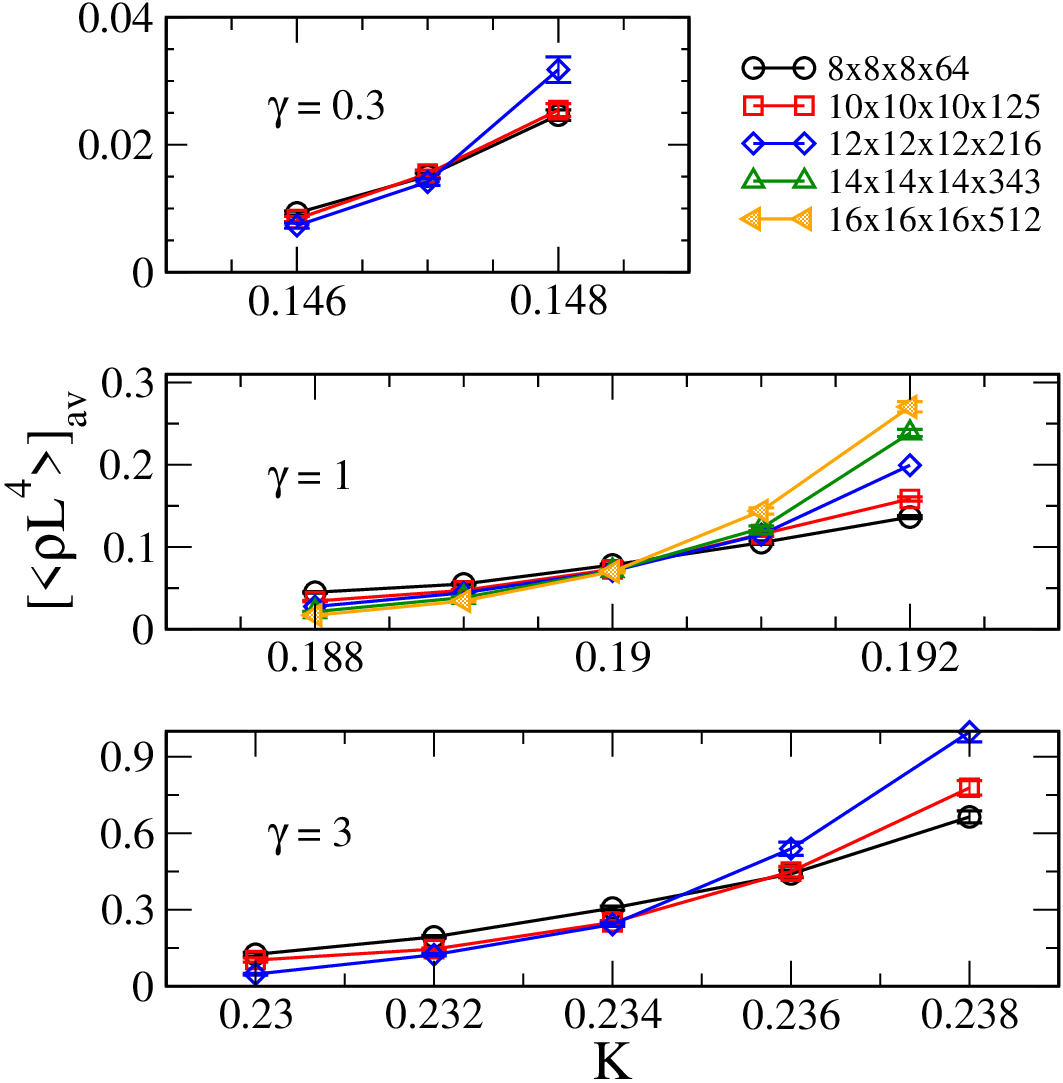}
\caption{ Crossings of $\bm{[\langle\rho L^4\rangle]_{av}}$ for three values of the anisotropy
				$\bm{\gamma} = $ 0.3, 1, and 3. The critical point increases as a function of $\bm{\gamma}$, as does the
				value of $\bm{[\langle\rho	L^4\rangle]_{av}}$ at the critical point.} 
\label{fig:gamma}
\end{figure}

We now outline how we arrive at these equivalent models. The transformation
from the Bose-Hubbard Hamiltonian to the effective link-current Hamiltonian described in
Ref.~\cite{wallin:1994:prb} yields, near the end of the procedure, an anisotropic action
\begin{eqnarray}
Z = \sum_{\bm{J}}\exp \bigg\{ -\sum_{({\bf r},\tau)}
\Big[\frac{K_x}{2}\bm{J'}^2_{({\bf r},\tau)} \nonumber\\
+ \frac{K_{\tau}}{2}{J^{\tau}_{({\bf r},\tau)}}^2 - \Delta \tau \bar{\mu_{{\bf r}}} J^{\tau}_{({\bf r},\tau)} \Big] \bigg\}
\end{eqnarray}
where $\Delta \tau$ is the width of each time slice, $\bm{J'} = \sum_{\nu = x, y, z}J^2$, and 
\begin{equation}
\label{eq:Kaniso}
K_t = U\Delta\tau,\qquad K_x = -2\ln (-t\Delta\tau /2).
\end{equation}
The simplest approach here is to set $KK_x = KK_\tau = 1$, which yields the isotropic link-current model as
stated above with $\mu_{{\bf r}} = \bar{\mu_{{\bf r}}} / U$.   There is a freedom here, however: one
can instead set $KK_x = 1$ and $KK_\tau = \gamma$, and introduce an anisotropy between the space and
time couplings in the link-current model without affecting the universal details:
\begin{equation} 
H = \frac{1}{K}\sum_{({\bf r}, \tau)}\bigg\{\frac{1}{2}\bm{J'}^2_{({\bf r},\tau)} 
+ \gamma \Big[\frac{1}{2}{J^{\tau}_{({\bf r}, \tau)}}^2 - \mu_{{\bf r}}J^\tau_{({\bf r}, \tau)}\Big]\bigg\}.  
\end{equation}

The link-current couplings can be mapped back to the Bose-Hubbard tunneling and on-site disorder
using (\ref{eq:Kaniso}):
\begin{equation}
\frac{U}{t} = 2\gamma\big(K e^{1/2K}\big)^{-1}.
\end{equation}
If the transition occurs at a particular ratio $U/t$, an increase in $\gamma$ implies then an
increase in $K_c$, for $K_c < 1/2$. More importantly, increasing $\gamma$ freezes out the temporal dynamics
of the link-current model. Both of these effects should speed up the spatial dynamics of the `worm'
algorithm, and since $\rho$ is defined in terms of these winding numbers, by changing $\gamma$ we can tune
the value of $\rho L^4$ at the critical point. Figure~\ref{fig:gamma} shows crossings in $\rho L^4$
for $\gamma = 0.3$, 1, and 3. As expected, increasing $\gamma$ increases the value of $K_c$ and the
magnitude of the stiffness at the crossing, hence improving the sampling efficiency of the algorithm. Note
that this has been achieved without changing the aspect ratio, $\alpha$.
 
Due to space constraints we do not show results for the critical exponents at different $\gamma$.
However, we have studied them in detail and they are indeed independent of $\gamma$ as one would
expect from universality arguments. We note that it would be of considerable interest to study the
evolution of the distribution of thermodynamic variable, as shown in Fig.~\ref{fig:equil} for
$\gamma=1$, for different values of $\gamma$. Preliminary results indicates that the width of the
distributions vary with $\gamma$, likely increasing monotonically with this parameter. Due to time
constraints we leave a detailed investigation for future study.

\Section{\label{sec:concl}Conclusions}
Numerical studies of disordered systems are notoriously difficult due to the large amount of
computational resources required and to the many subtle sources of systematic error. The procedure
we have presented to estimate disorder distributions and their averages highlights some of the
potential pitfalls. Under-sampling the disorder distribution (setting $N_D$ too low), under-sampling
each individual distribution (setting $t_s$ too low), and failing to properly equilibrate each
simulation (setting $t_0$ too low) can all lead to erroneous estimates. The two procedures presented
above provide confirmation that these pitfalls have been avoided. The convergence of the Hamming
distances provides a measure of $t_r$ and confirms that the Monte Carlo simulations are generating
equilibrium configurations. The disorder distributions themselves can contain artifacts of the Monte
Carlo averaging procedure; they should not demonstrate any dependence on $t_0$ or $t_s$ if these
have been set sufficiently large. These procedures can be generalized to other systems.

In the context of the Bose-Hubbard model, the discrete nature of the winding number in the
link-current representation forces the investment of a large amount of computational resources in
simulating each individual disorder realization ($t_s$ must be set much greater than $t_r$).
However, by adjusting the anisotropy between the spatial and temporal coupling strengths in the
link-current model, we can increase the magnitude of the superfluid stiffness at the critical point.
This improves the numerical efficiency of the calculation without affecting the universal details of
the critical behaviour. This observation may prove useful for future Monte Carlo studies of the
Bose-Hubbard model in the link-current representation.

\section*{Acknowledgements}
This work was supported by the Natural Sciences and Engineering Research Council of Canada, and by
SHARCNET. Computation was carried out on SHARCNET clusters at McMaster University. 

\bibliographystyle{latex8}
\bibliography{bhd3d}
\end{document}